\documentclass[11pt]{article}
\usepackage{amssymb,amsfonts,cite,amsmath,graphicx,epsf}
\usepackage{epstopdf}
\usepackage{psfrag}
\usepackage{bbm}

\usepackage[totalwidth=151mm,totalheight=245mm,centering]{geometry}

\newcommand\blank[1]{}

\newcommand{\fract}[2]{{\textstyle\frac{#1}{#2}}}

\newcommand{\ep}{\epsilon}

\newcommand\ZZ{{\mathbb Z}}
\newcommand\RR{{\mathbb R}}

\newcommand\phup{^{\phantom o}}

\newcommand{\CM}{{\cal M}}

\newcommand\eq{\begin{equation}}
\newcommand\en{\end{equation}}
\newcommand\bea{\begin{eqnarray}}
\newcommand\eea{\end{eqnarray}}
\newcommand\nn{\nonumber}

\newcommand\MR{M\!R}

\newcommand{\vev}[1]{\langle #1 \rangle}
\newcommand{\ket}[1]{|#1\rangle}
\newcommand{\iket}[1]{|\,#1\,\rangle\!\rangle}

\newcommand{\resection}[1]{\setcounter{equation}{0}\section{#1}}

\begin{document}

\begin{titlepage} 
\vskip 0.5cm
\begin{flushright}
DCPT--09/81  \\
\end{flushright}
\vskip 1.8cm
\begin{center}
{\Large\bf Exact g-function flow between conformal field theories
} \\[5pt]
\end{center}
\vskip 0.8cm

\centerline{Patrick Dorey%
\footnote{{\tt p.e.dorey@durham.ac.uk}},
Chaiho Rim
\footnote{{\tt rimpine@sogang.ac.kr}}
and Roberto Tateo%
\footnote{{\tt tateo@to.infn.it}}}
\vskip 0.9cm
\centerline{${}^{1}$\sl\small Department of Mathematical Sciences,
Durham University,}
\centerline{\sl\small South Road, Durham DH1 3LE, UK}
\vskip 0.3cm 
\centerline{${}^{2}$\sl\small Department of Physics and CQUEST, 
Sogang University,}
\centerline{\sl\small Seoul, 121-742, Republic of Korea}
\vskip 0.3cm 
\centerline{${}^{3}$\sl\small Dip.\ di Fisica Teorica
and INFN, Universit\`a di Torino,} 
\centerline{\sl\small Via P.\ Giuria 1, 10125 Torino, Italy}
\vskip 1.25cm
\vskip 0.9cm
\begin{abstract}
\vskip0.15cm
\noindent
Exact equations are proposed to describe $g$-function flows
in integrable boundary quantum field theories which interpolate between 
different conformal field theories in their ultraviolet and infrared 
limits, extending previous work where purely massive flows were
treated. The approach is illustrated with flows between the 
tricritical and critical Ising models, but the method is not
restricted to these cases and should be of use in unravelling general 
patterns of integrable boundary
flows between pairs of conformal field theories.
\end{abstract}
\end{titlepage}
\setcounter{footnote}{0}

\resection{Introduction}
Since work by A.B.~Zamolodchikov more than 20 years 
ago \cite{Zamolodchikov:1987ti}, many examples of
two-dimensional quantum field theories which flow between 
different conformal field theories in their short and long distance
limits
have been found. If such theories are placed on manifolds with
one or more boundaries, then the corresponding boundary conditions
must also flow, between conformal boundary conditions appropriate to
the conformal field theories sitting at the two limits.
To figure out the resulting pattern of combined bulk and boundary
flows is an interesting problem, with potential
relevance to a variety of issues in condensed matter physics and 
string theory.

Zamolodchikov's original paper (see also \cite{Ludwig:1987gs})
concerned the bulk perturbation of the unitary minimal model
$\CM_{p,p+1}$ by its $\phi_{13}$ operator. For $p$ large, a 
perturbative calculation of the $c$-function 
\cite{Zamolodchikov:1986gt} enabled him to show that,
for one sign of the coupling, 
the resulting renormalisation group flow
interpolates between $\CM_{p,p+1}$ in the ultraviolet
and $\CM_{p-1,p}$ in the infrared. The generalisation of
this approach to the boundary situation is surprisingly tricky, but
has recently been achieved in a paper by Fredenhagen, Gaberdiel and
Schmidt-Colinet \cite{Fredenhagen:2009tn}, where the $g$-function
or boundary entropy \cite{Tsvelick:1985,Affleck:1991tk}
was used to identify the
destination boundary conditions, again for large $p$.
(Even more recently, the same pattern of flows has been shown to hold 
on fluctuating surfaces with boundaries \cite{Bourgine:2009zt}.)

The calculations of \cite{Fredenhagen:2009tn}
are perturbative in $1/p$, and they do not give
reliable information about flows near the bottom of the minimal series.
In fact, even at large $p$ the authors of~\cite{Fredenhagen:2009tn}
had to borrow some
non-perturbative information about pure-boundary flows in order to
obtain a full picture.  In the absence of
boundaries, Al.B.~Zamolodchikov showed how such problems could be
circumvented in integrable situations through the use of exact,
nonperturbative equations of 
Thermodynamic Bethe Ansatz (TBA) type~\cite{Zamolodchikov:1991vx}.
These equations encode the evolution of a quantity called the
effective central charge,
$c_{\rm eff}$, during renormalisation group
flows, where
$c_{\rm eff}$ is an off-critical generalisation of the central 
charge of a conformal field
theory, which agrees with the $c$-function used in
\cite{Zamolodchikov:1987ti} at
fixed points of the renormalisation group.
The purpose of this paper is to show that a similarly-exact description 
of bulk flows with boundaries is possible, at least in cases where the
combined bulk and boundary theory is integrable. Our starting-point is 
the exact off-critical $g$-function for massive integrable quantum
field theories that was
introduced in \cite{Dorey:2004xk}
and further studied in \cite{Dorey:2005ak}. After some
background material in section \ref{background}, the proposed massless
variant of the TBA-inspired
exact $g$-function of \cite{Dorey:2004xk} is introduced 
in section \ref{proposal}, together with some numerical illustrations of
its implications. These results are backed by exact
calculations of limiting $g$-function values in section \ref{checks},
where we also report some simple perturbative checks of our proposal.
Finally section \ref{conclusions} contains some conclusions.

In cases where the bulk remains critical, the use of equations of TBA
type to evaluate $g$-functions has a long history, dating back at
least to work on the Kondo problem~\cite{Tsvelick:1985}. In this
respect the main novelty of our result is the demonstration that,
for off-critical interpolating flows,
bulk-induced changes to $g$-functions 
can also be accounted for, exactly, through the TBA approach.
Some motivation for our specific proposal came from a consideration of
Al.B.~Zamolodchikov's
staircase model~\cite{Zamolodchikov:1991pc} (see also
\cite{Martins:1992ht,Dorey:1992bq,Martins:1992sx,Dorey:1992pj}).
The full set of flows implied by this connection is rather
rich, and we postpone its discussion to another
occasion\footnote{Though we should mention a previous attempt to use 
the staircase TBA to study bulk- and boundary- induced $g$-function flows 
between conformal field theories, reported in
\cite{Lesage:1998qf}.
However the equations developed there do not fully account for the
effects of an off-critical bulk on the $g$-functions, and do not agree
with our results.}.
We have also limited the treatment in this paper to flows between 
the tricritical Ising and Ising models (noting, 
though, that these cases are of particular
interest, being the furthest
possible from the perturbative limit studied previously).
Generalisation to other cases appears to be straightforward, and we
plan to present a more detailed analysis elsewhere.

\resection{Background}
\label{background}

\subsection{The bulk flow}
The TBA system 
found in \cite{Zamolodchikov:1991vx} 
for the tricritical Ising to Ising flow 
encodes the ground state energy $E(R)$ of the interpolating
theory on a circle of circumference $R$ via a pair 
of integral equations for two pseudoenergies $\ep_1(\theta)$ and
$\ep_2(\theta)$. Using the symmetry $\ep_2(\theta)=\ep_1(-\theta)$
these boil down to a single equation,
for $\ep(\theta)\equiv\ep_1(\theta)=\ep_2(-\theta)$\,:
\eq
\ep(\theta)=\frac{1}{2}\,re^{\theta}-
\int_{\RR}\phi(\theta+\theta')\,L(\theta')\,d\theta'\,.
\label{bulkTBA}
\en
Here $L(\theta)=\ln(1+e^{-\ep(\theta)})$,
$\phi(\theta)=\frac{1}{2\pi\cosh(\theta)}$\,,
and $r=\MR$ with $M$ a parameter with the dimensions of mass 
which
sets the (inverse) crossover scale. Then 
\eq
E(R)= -\frac{\pi}{6R}\,c_{\rm eff}(r)
\en
where 
\eq
c_{\rm eff}(r)=\frac{3}{\pi^2}\,\int_{\RR}
re^{\theta}L(\theta)\,d\theta\,.
\label{ceff}
\en
The limiting values $c_{\rm eff}(0)=7/10$ and 
$c_{\rm eff}(\infty)=1/2$ can be calculated 
exactly~\cite{Zamolodchikov:1991vx}, and
match the central charges of the tricritical Ising and Ising models.
Later, the form of $L(\theta)$ in these two limits
will be important. As $r\to 0$, three regions develop where
$L(\theta)$ is approximately constant: 
\begin{align}
&\quad&L(\theta)&\sim \ln(2) =0.6931\dots
 &\text{for}~~& \theta \ll -\ln(1/r) ~;\label{UVa}\\
&\quad&L(\theta)&\sim \ln((3{+}\sqrt{5})/2)=0.9624\dots
\!\!\!\!\!\!\!\!
 &\text{for}~~& -\ln(1/r)\ll\theta\ll\ln(1/r) ~;~ \label{UVb}\\[2pt]
&\quad&L(\theta)&\sim 0
 &\text{for}~~& \theta\gg\ln(1/r) ~.\label{UVc}
\end{align}
In the opposite limit, $r\to\infty$, there are instead just two regions:
\begin{align}
&\quad&L(\theta)&\sim \ln(2) 
 &\text{for}~~& \theta \ll -\ln(r) ~;&\qquad& \label{IRa}\\
&\quad&L(\theta)&\sim 0
 &\text{for}~~& \theta\gg -\ln(r) ~.&\qquad& \label{IRb}
\end{align}
These behaviours are illustrated in figure \ref{Lvals}.

\[
\begin{array}{c}
\includegraphics[width=0.95\linewidth]{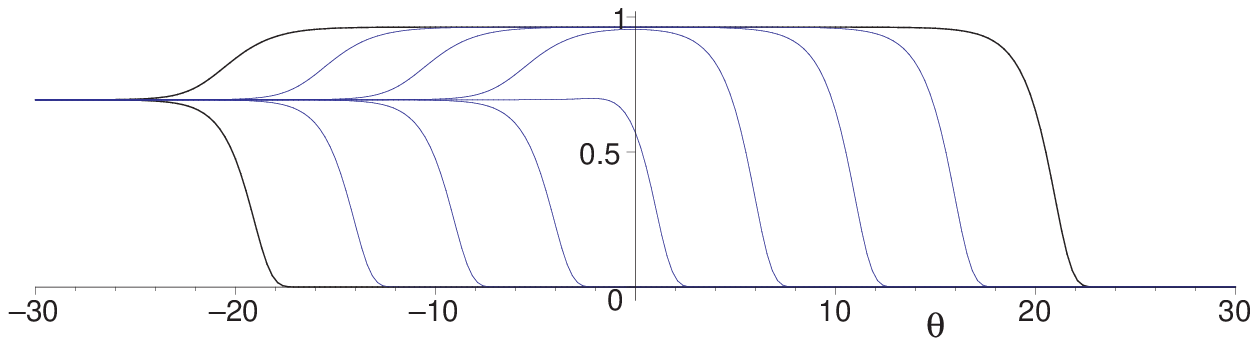}
\\[11pt]
\parbox{0.78\linewidth}{
\small Figure \protect\ref{Lvals}: 
$L(\theta)$ for various values of $r$. From the uppermost to
the lowermost curve, the values of $\ln(r)$ run from $-20$ to
$20$ in equal steps.
}
\end{array}
\]
\refstepcounter{figure}
\protect\label{Lvals}

\subsection{The conformal boundary conditions}
With the addition of boundaries, the endpoints of the interpolating
flows become
boundary conformal field theories: the boundary tricritical Ising 
model in the ultraviolet, and the boundary (critical) Ising model in 
the infrared. The basic (Cardy) boundary states for these two models 
follow from \cite{Cardy:1989ir}, and their physical interpretations
were discussed in \cite{Cardy:1989ir} (for Ising) and
\cite{Chim:1995kf,Affleck:2000jv} (for tricritical Ising). For Ising,
there are three possibilities, corresponding to the boundary spins
being fixed up $(+)$, fixed
down ($-$), or free ($f$). Written in terms of Ishibashi
states~\cite{Ishibashi:1988kg} 
$\iket{0}$, $\iket{\varepsilon}$ and $\iket{\sigma}$ 
they are
\cite{Cardy:1989ir}
\begin{align}
\ket{(+)}&=
\frac{1}{\sqrt{2}}\iket{0}+\frac{1}{\sqrt{2}}\iket{\varepsilon}
+\frac{1}{\sqrt[4]{2}}\iket{\sigma}\nn\\[2pt]
\ket{(-)}&=
\frac{1}{\sqrt{2}}\iket{0}+\frac{1}{\sqrt{2}}\iket{\varepsilon}
-\frac{1}{\sqrt[4]{2}}\iket{\sigma}\nn\\[4pt]
\ket{(f)}&=\iket{0}-\iket{\varepsilon}
\label{Isingbndry}
\end{align}
The inner products of these states with the vacuum $\ket{0}$
give the corresponding values of the conformal
$g$-function \cite{Affleck:1991tk}. On Ishibashi states
$\iket{\alpha}$ we have
$\langle 0\,\iket{\alpha}=\delta_{0\alpha}$, so
\begin{align}
\ln g\phup_{(+)}\big|_{\rm Ising} =\ln g\phup_{(-)}\big|_{\rm Ising} 
= \ln\frac{1}{\sqrt{2}}&=-0.3465\dots\\[3pt]
\ln g\phup_{(f)}\big|_{\rm Ising}=\ln 1&=\phantom{-}0\,.
\end{align}
We will also treat the superposition of $(+)$ and
$(-)$ boundaries, $(+)\&(-)$, for which
\eq
\ln g\phup_{(+)\&(-)}\big|_{\rm Ising}=
\ln 2g\phup_{(+)}\big|_{\rm Ising}=\phantom{-}0.3465\dots
\en

\smallskip

For the tricritical Ising model there are instead six options, each
labelled, roughly speaking, by the value (or values) available to the
order parameter $\vev{\sigma}$ at that boundary, taken from 
$\{-,0,+\}$ \cite{Chim:1995kf}. (In the conformal field theory, $\sigma$ 
becomes the leading spin
field, with dimensions $(3/80,3/80)$.)
These are $(-)$, $(0)$, $(+)$, $(-0)$, 
$(0+)$, and $(-0+)$, though the last of these is traditionally labelled
as $(d)$, with $d$ standing for `degenerate'. The corresponding
boundary states are\,\footnote{Note, our
assignments of the $\ket{(+)}$ and $\ket{(-)}$ states, and of the
$\ket{(0+)}$ and $\ket{(-0)}$ states, are opposite to those in
\cite{Chim:1995kf}. This is to ensure that the one-point
function of the spin field $\sigma$ is positive in the presence of
the $(+)$ boundary, and negative in the presence of the $(-)$
boundary, which is more natural, and matches the
convention adopted in (\ref{Isingbndry}) for the 
Ising boundary states. We've also corrected a typo in the $\ket{(+)}$
and $\ket{(-)}$ boundary states as given in \cite{Chim:1995kf}; our
states match those given in, for example, \cite{Nepomechie:2001bu}.}
\begin{align}
\ket{(+)}~&=
C\Bigl[\,
\iket{0} +\eta\iket{\fract{1}{10}} +\eta\iket{\fract{3}{5}}
+\iket{\fract{3}{2}} +\sqrt[4]{2}\iket{\fract{7}{16}}
+\sqrt[4]{2}\eta\iket{\fract{3}{80}} \,\Bigr]\nn\\[2pt]
\ket{(-)}~&=
C\Bigl[\,
\iket{0} +\eta\iket{\fract{1}{10}} +\eta\iket{\fract{3}{5}}
+\iket{\fract{3}{2}} -\sqrt[4]{2}\iket{\fract{7}{16}}
-\sqrt[4]{2}\eta\iket{\fract{3}{80}} \,\Bigr]\nn\\[2pt]
\ket{(0)}~\,&=
\sqrt{2}C\Bigl[\,
\iket{0} -\eta\iket{\fract{1}{10}} +\eta\iket{\fract{3}{5}}
-\iket{\fract{3}{2}} \,\Bigr]\nn\\[2pt]
\ket{(0+)}&=
C\Bigl[\,
\eta^2\iket{0} -\eta^{-1}\iket{\fract{1}{10}} -\eta^{-1}\iket{\fract{3}{5}}
+\eta^2\iket{\fract{3}{2}} -\sqrt[4]{2}\eta^2\iket{\fract{7}{16}}
+\sqrt[4]{2}\eta^{-1}\iket{\fract{3}{80}} \,\Bigr]\nn\\[2pt]
\ket{(-0)}&=
C\Bigl[\,
\eta^2\iket{0} -\eta^{-1}\iket{\fract{1}{10}} -\eta^{-1}\iket{\fract{3}{5}}
+\eta^2\iket{\fract{3}{2}} +\sqrt[4]{2}\eta^2\iket{\fract{7}{16}}
-\sqrt[4]{2}\eta^{-1}\iket{\fract{3}{80}} \,\Bigr]\nn\\[2pt]
\ket{(d)}~\,&=
\sqrt{2}C\Bigl[\,
\eta^2\iket{0} +\eta^{-1}\iket{\fract{1}{10}} -\eta^{-1}\iket{\fract{3}{5}}
-\eta^2\iket{\fract{3}{2}}  \,\Bigr]
\label{TIMstates}
\end{align}
where 
\eq
C=\sqrt{\sin(\pi/5)/\sqrt{5}}=
\left(\frac{1}{8}-\frac{1}{8\sqrt{5}}\right)^{1/4}~;\quad
\eta=\sqrt{2\cos(\pi/5)}=\sqrt{(1{+}\sqrt{5})/2}~.
\en
Again adding in the $(+)\&(-)$ superposition, the $g$-function values 
we will need, ordered by increasing $g$ and
expressed in ways that will be useful for comparisons
later on, are
\begin{align}
\ln g\phup_{(+)}\big|_{\rm tricrit} =\ln g\phup_{(-)}\big|_{\rm
tricrit}  = \ln(C)
=\fract{1}{4}\ln(\fract{1}{8}{-}\fract{1}{8\sqrt{5}})
&=-0.6680\dots\\[3pt]
\ln g\phup_{(0)}\big|_{\rm tricrit} 
  = \ln(\sqrt{2}C)
=\fract{1}{4}\ln(\fract{1}{2}{-}\fract{1}{2\sqrt{5}})
&=-0.3214\dots\\[3pt]
\ln g\phup_{(0+)}\big|_{\rm tricrit} =\ln g\phup_{(-0)}\big|_{\rm
tricrit}  = \ln(\eta^2C)
=\fract{1}{4}\ln(\fract{1}{4}{+}\fract{1}{2\sqrt{5}})
&=-0.1868\dots\\[3pt]
\ln g\phup_{(+)\&(-)}\big|_{\rm tricrit}=
\ln 2g\phup_{(+)}\big|_{\rm tricrit} =
\ln g\phup_{(0)}\big|_{\rm tricrit}+\fract{1}{2}\ln 2
&=\phantom{-}0.0250\dots
\\[3pt]
\ln g\phup_{(d)}\big|_{\rm tricrit} 
  = \ln(\sqrt{2}\eta^2C)
= \ln g\phup_{(0+)}\big|_{\rm tricrit}+\fract{1}{2}\ln 2
&=\phantom{-}0.1597\dots
\end{align}

\subsection{The boundary flows}
The flows which occur when the two models are perturbed at the
boundary alone are well-understood. In Ising, the $(f)$ boundary
admits a single relevant boundary field $\phi_{13}$, with dimension
$1/2$. This
breaks the $\ZZ_2$ symmetry of the bulk and
can be interpreted as a boundary magnetic field. Depending on the sign
of the perturbation, a flow is induced to the $(-)$ or to the $(+)$
boundary, as shown in figure \ref{IMflows1}.
\[
\begin{array}{c}
\includegraphics[height=0.04\linewidth]{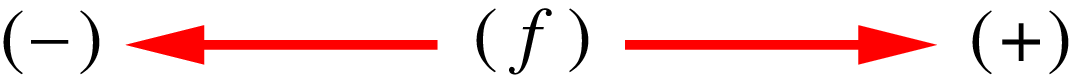}
\\[11pt]
\parbox{0.6\linewidth}{
\small Figure \protect\ref{IMflows1}: 
Flows from the $(f)$ boundary in the Ising model.
}
\end{array}
\]
\refstepcounter{figure}\protect\label{IMflows1}%
The $(+)$ and $(-)$ boundaries have no relevant boundary fields, but
one can also consider their superposition,
$(+)\&(-)$. The boundary-condition changing
operators correspond to $\phi_{13}$ again, and generate the flow
illustrated in figure \ref{IMflows2}, to
the free boundary condition \cite{COT,Watts:2000kp}:
\[
\begin{array}{c}
\includegraphics[height=0.04\linewidth]{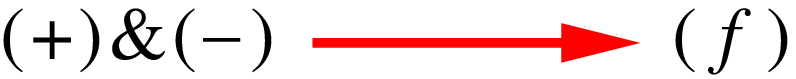}
\\[11pt]
\parbox{0.67\linewidth}{
\small Figure \protect\ref{IMflows2}: 
The flow from the $(+)\&(-)$ boundary in the Ising model.
}
\end{array}
\]
\refstepcounter{figure}\protect\label{IMflows2}%

For the tricritical Ising model the structure is richer
\cite{Chim:1995kf,Affleck:2000jv} (see also 
\cite{Graham:2000si,Feverati:2003rb,Nepomechie:2002ak}). 
Including the superposition $(+)\&(-)$, the
full map is shown in figure \ref{TIMflows}.
\[
\begin{array}{c}
\includegraphics[height=0.37\linewidth]{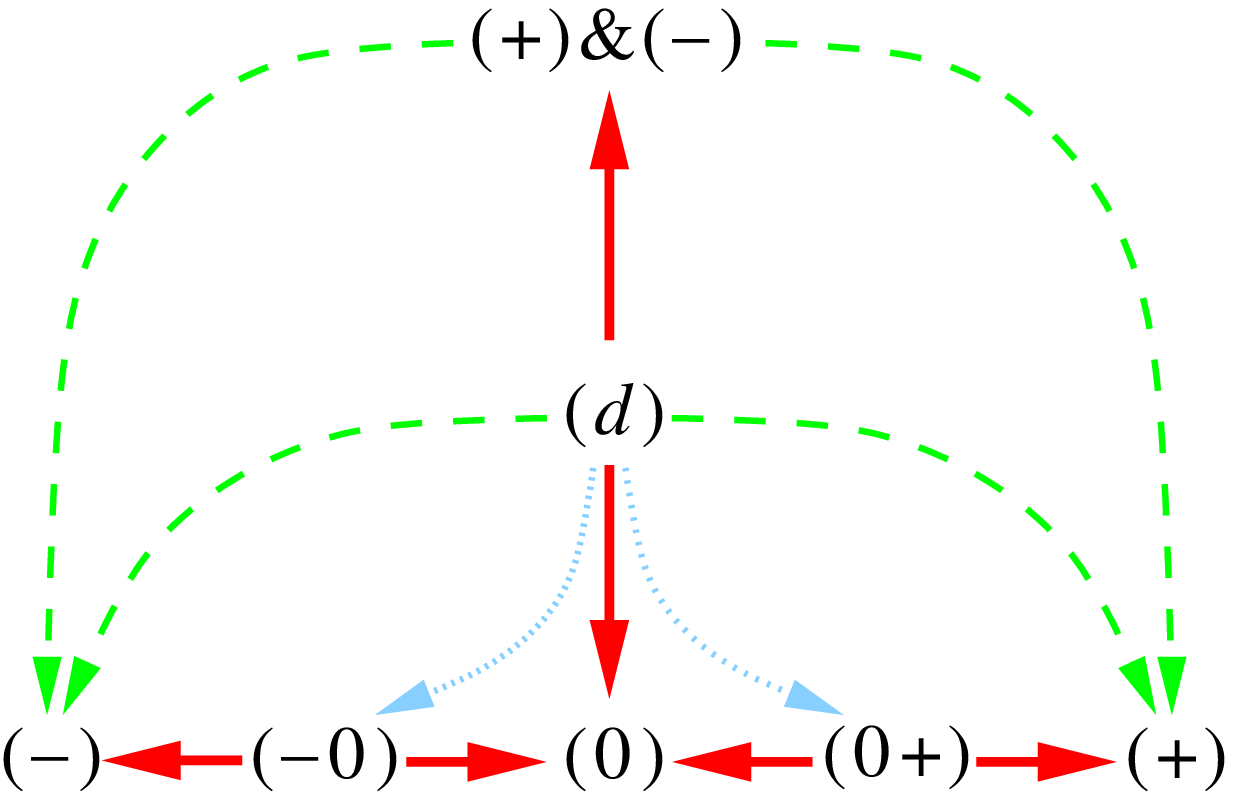}
\\[14pt]
\parbox{0.8\linewidth}{
\small Figure \protect\ref{TIMflows}: 
Boundary flows in the tricritical Ising model. Solid lines (red
online) show flows induced by a $\phi_{13}$ boundary field.
The dashed lines (green online) are induced by $\phi_{12}$ (for the
lower two lines) or $\phi_{11}$ (for the upper two). Finally,
the dotted
lines (blue online) are induced by a combination of $\phi_{12}$
and $\phi_{13}$.
}
\end{array}
\]
\refstepcounter{figure}\protect\label{TIMflows}%

With the bulk conformal, the flows induced by $\phi_{13}$, 
$\phi_{12}$
and $\phi_{11}$ (the solid and dashed lines on figure \ref{TIMflows})
are all integrable. However, for integrability to survive when the
bulk is also perturbed, it is not enough for the bulk and boundary 
perturbations to be separately integrable -- they must also be
compatible with each other. For the $\phi_{13}$ bulk perturbation
which leads to the interpolating flow to Ising, this is thought to 
hold if the boundary perturbing operator is also $\phi_{13}$
\cite{Ghoshal:1993tm}, and so it is these combined bulk-and-boundary
flows that we should aim to treat using the exact $g$-function.

\resection{Exact g-functions for the interpolating flow}
\label{proposal}
In \cite{Dorey:2004xk,Dorey:2005ak}, exact equations were proposed
for the off-critical $g$-function in certain massive integrable 
boundary theories. To give our proposal for the interpolating
bulk and boundary
tricritical Ising to Ising flows, we first set
\eq
(x)(\theta)=
\frac%
{\sinh\left(\fract{\theta}{2}+\fract{i\pi x}{2}\right)}%
{\sinh\left(\fract{\theta}{2}-\fract{i\pi x}{2}\right)}~,\qquad
\phi_{(x)}(\theta)=-\frac{i}{2\pi}\frac{d}{d\theta}\ln\,(x)(\theta)=
\frac{-\sin(\pi x)/(2\pi)\,}{\cosh(\theta)-\cos(\pi x)}~,
\en
so that the kernel function $\phi(\theta)$ in the bulk TBA equation 
(\ref{bulkTBA}) is equal to $-\phi_{(1/2)}(\theta)$,
and 
\eq
\int_{\RR}\phi_{(x)}(\theta)\,d\theta=-(1-|x|)\,{\rm sgn}(x)\,.
\label{intform}
\en
Now let $\ep(\theta)$ solve the bulk TBA equation 
(\ref{bulkTBA}) for a system on a cylinder of circumference $r$, and
suppose a boundary is placed at the end of that cylinder with a
boundary condition which depends on a further parameter $\theta_b$.
We will propose the following expression for the logarithm of a
$g$-function $\ln g(r)$:
\eq
\ln g(r)=
\ln g\phup_0(r)+ \ln g_b(r)
\label{gformula}
\en
where
\eq
\ln g\phup_0(r)=
\sum_{j=1}^{\infty}
\frac{1}{2j{-}1}
\int_{\RR^{2j-1}} \frac{d \theta_1}{1+e^{\ep(\theta_1)}} \dots
\frac{d \theta_{2j{-}1} }%
{ 1+e^{\ep(\theta_{2j{-}1})}}\,
\phi(\theta_1+\theta_2)\phi(\theta_2+\theta_3)\dots
\phi(\theta_{2j{-}1}+\theta_{1})
\label{g0}
\en
and
\eq
\ln g_b(r)= -\frac{1}{2}\ln(2)+ \int_{\RR} \,
 (\phi_b(\theta)  - \phi(2 \theta) )
 L(\theta)\, d \theta 
\label{gb}
\en
with $L(\theta)=\ln(1+e^{-\ep(\theta)})$ as before, and
\eq
\phi_{b}(\theta)=\phi_{(3/4)}(\theta)-\phi_{(1/2)}(\theta-\theta_b)\,.
\en
(Note, the normalisation of $\phi_b$ here  differs
by a factor of $2$ from that in~\cite{Dorey:2004xk,Dorey:2005ak}.)

The expression (\ref{gformula})
has the same general structure as the exact massive
$g$-function introduced in 
\cite{Dorey:2004xk}, with $g_b(r)$ containing the
boundary-condition specific parts of the $g$-function, while $g\phup_0(r)$
is a `universal' piece which incorporates the effects of the bulk
perturbation on the boundary entropy. However, the new formula
involves some significant changes too -- in particular,
the infinite series in (\ref{g0}) contains only odd terms, and all
rapidity combinations in the kernel functions
$\phi(\theta_i{+}\theta_{i{+}1})$ appear as sums.
(This second aspect is related to the fact that
(\ref{g0}) has been written in terms of the single function
$\ep(\theta)=\ep_1(\theta)$, rather than $\ep_1(\theta)$ and
$\ep_2(\theta)=\ep_1(-\theta)$.)

The infinite series gives an expansion for $\ln g_0(r)$ about
$r=\infty$, but it converges rapidly for all values of $r$, and
can be summed exactly at $r=0$, and in various intermediate
double-scaling limits. Before giving these
details, figures \ref{gflowm15}, \ref{gflowp00} and \ref{gflowp15}
show numerically-obtained plots of $\ln g(r)$ for
$\theta_b=-15$, $0$ and $+15$. The plots were obtained using 
the first five terms from the series (\ref{g0}), though
truncating to just three terms would have given 
visually indistinguishable results.
For all three values of $\theta_b$,
$\ln g(r)$ tends to $\ln g\phup_{(0+)}\big|_{\rm tricrit}$
in the far ultraviolet, to $\ln g\phup_{(+)}\big|_{\rm Ising}$
in the far infrared\footnote{Note, though, that these are equally the 
values of $\ln g\phup_{(0-)}\big|_{\rm tricrit}$ and $\ln
g\phup_{(-)}\big|_{\rm Ising}$ respectively.
We will mostly leave this ambiguity implicit in the following, but we
will return to it briefly later in this section.}, and undergoes
a transition  at $\ln r \approx 0$, which is where the
bulk crossover occurs. For $\theta_b=-15$, there are two further
transitions, at $\ln r\approx \pm 15$, while for $\theta_b=+15$ there 
is one, at $\ln r\approx -15$.

{~}\vspace{-30pt}
\[
\begin{array}{c}
\includegraphics[width=1.0\linewidth]{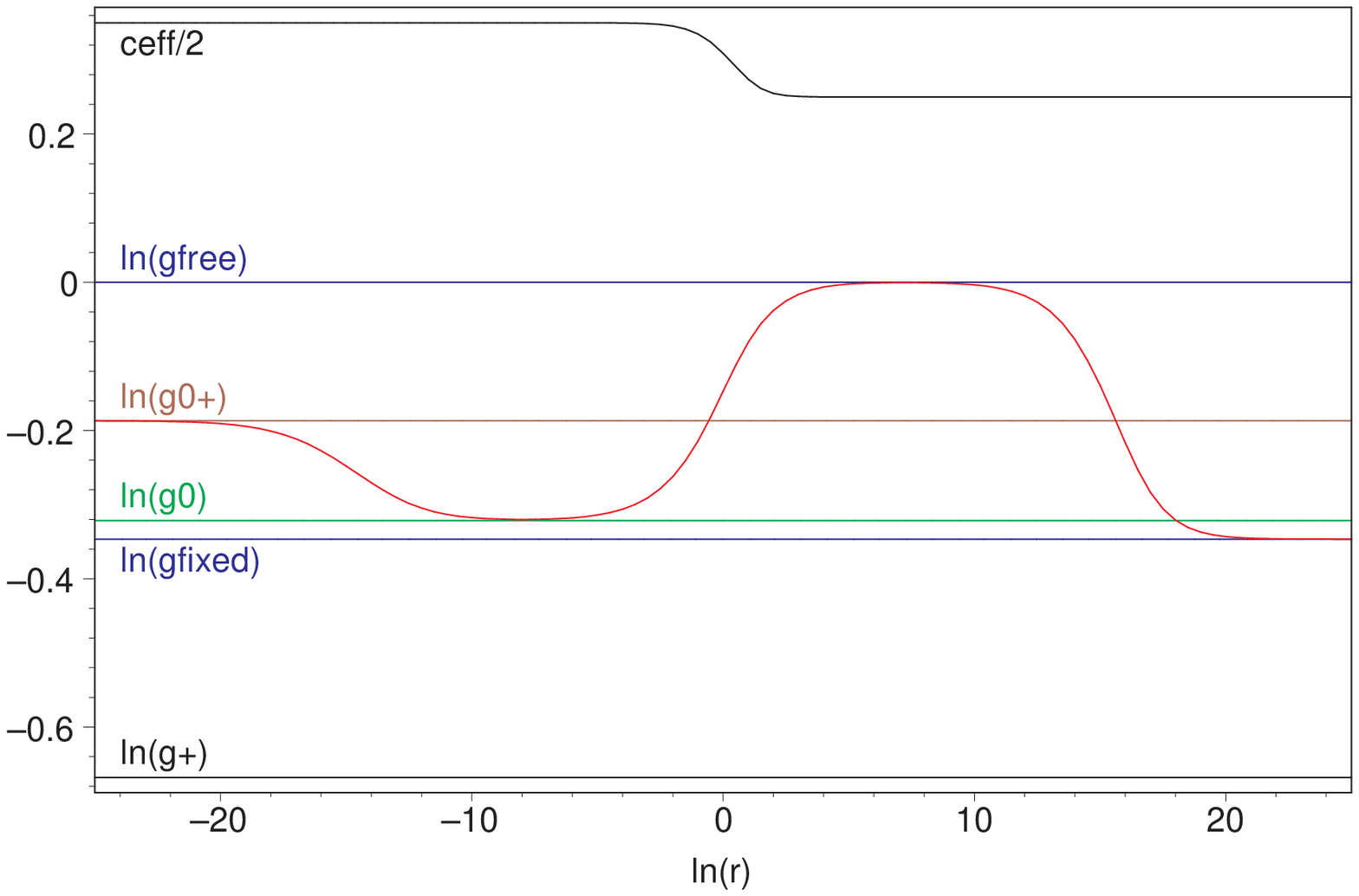}
\\[2pt]
\parbox{0.8\linewidth}{
\small Figure \protect\ref{gflowm15}: 
The exact $g$-function flow for $\theta_b=-15$\,. The
flow of $c_{\rm eff}(r)/2$, running from $0.35$ down to $0.25$,
is also shown, to indicate the location and duration of
the bulk crossover. Tricritcal Ising $g$-function values are 
{\fontsize{9.5}{0}\usefont{T1}{phv}{m}{n}g0+},
{\fontsize{9.5}{0}\usefont{T1}{phv}{m}{n}g0} and 
{\fontsize{9.5}{0}\usefont{T1}{phv}{m}{n}g+};
critical Ising values are
{\fontsize{9.5}{0}\usefont{T1}{phv}{m}{n} gfree} and
{\fontsize{9.5}{0}\usefont{T1}{phv}{m}{n}gfixed}.
}
\end{array}
\]
\refstepcounter{figure}\protect\label{gflowm15}%

{~}\vspace{-36pt}
\[
\begin{array}{c}
\includegraphics[width=1.0\linewidth]{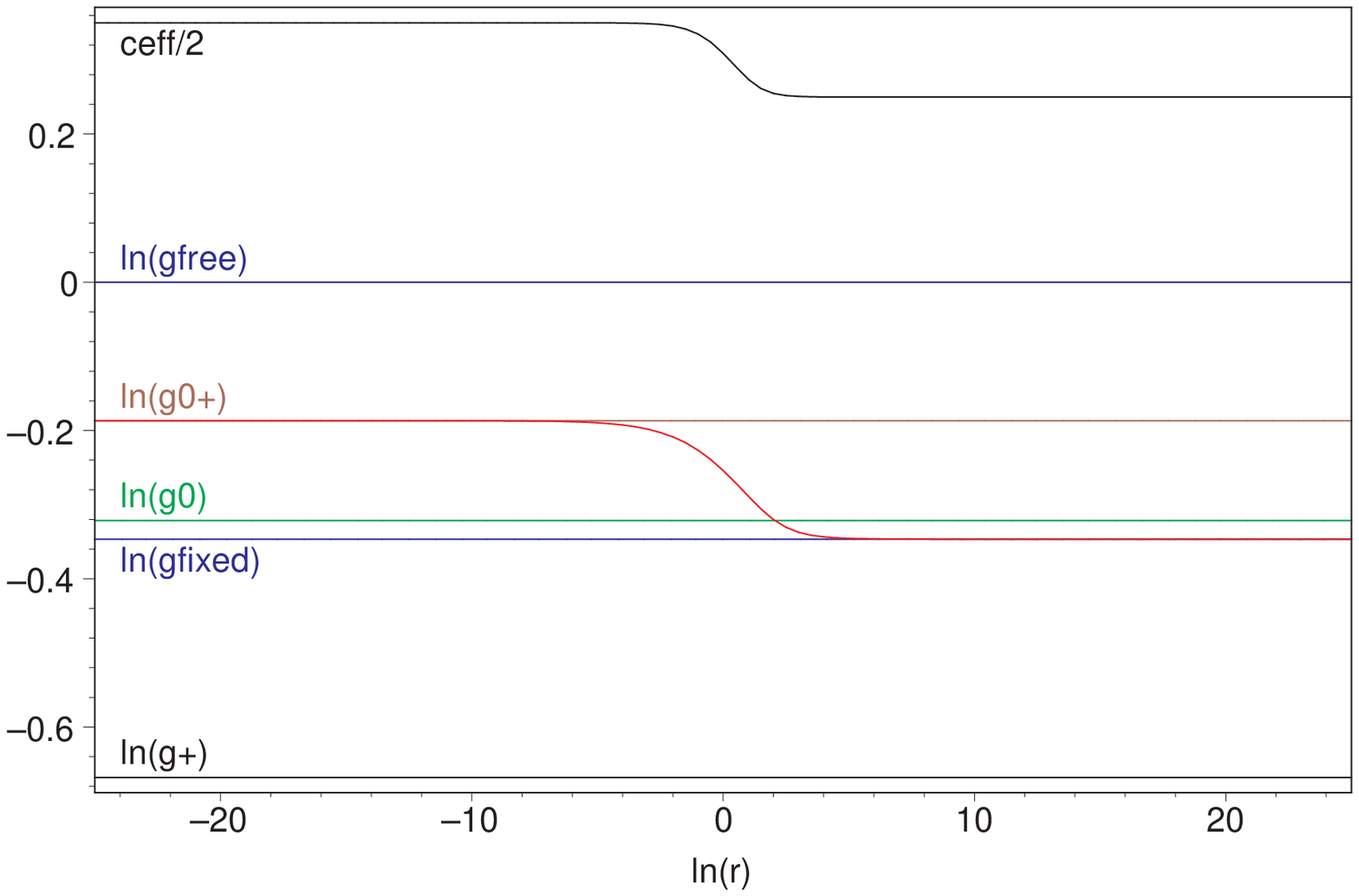}
\\[2pt]
\parbox{0.8\linewidth}{
\small Figure \protect\ref{gflowp00}: 
The exact $g$-function flow for $\theta_b=0$\,. Labelling as
for figure \ref{gflowm15}.
}
\end{array}
\]
\refstepcounter{figure}\protect\label{gflowp00}%

{~}\vspace{-35pt}
\[
\begin{array}{c}
\includegraphics[width=1.0\linewidth]{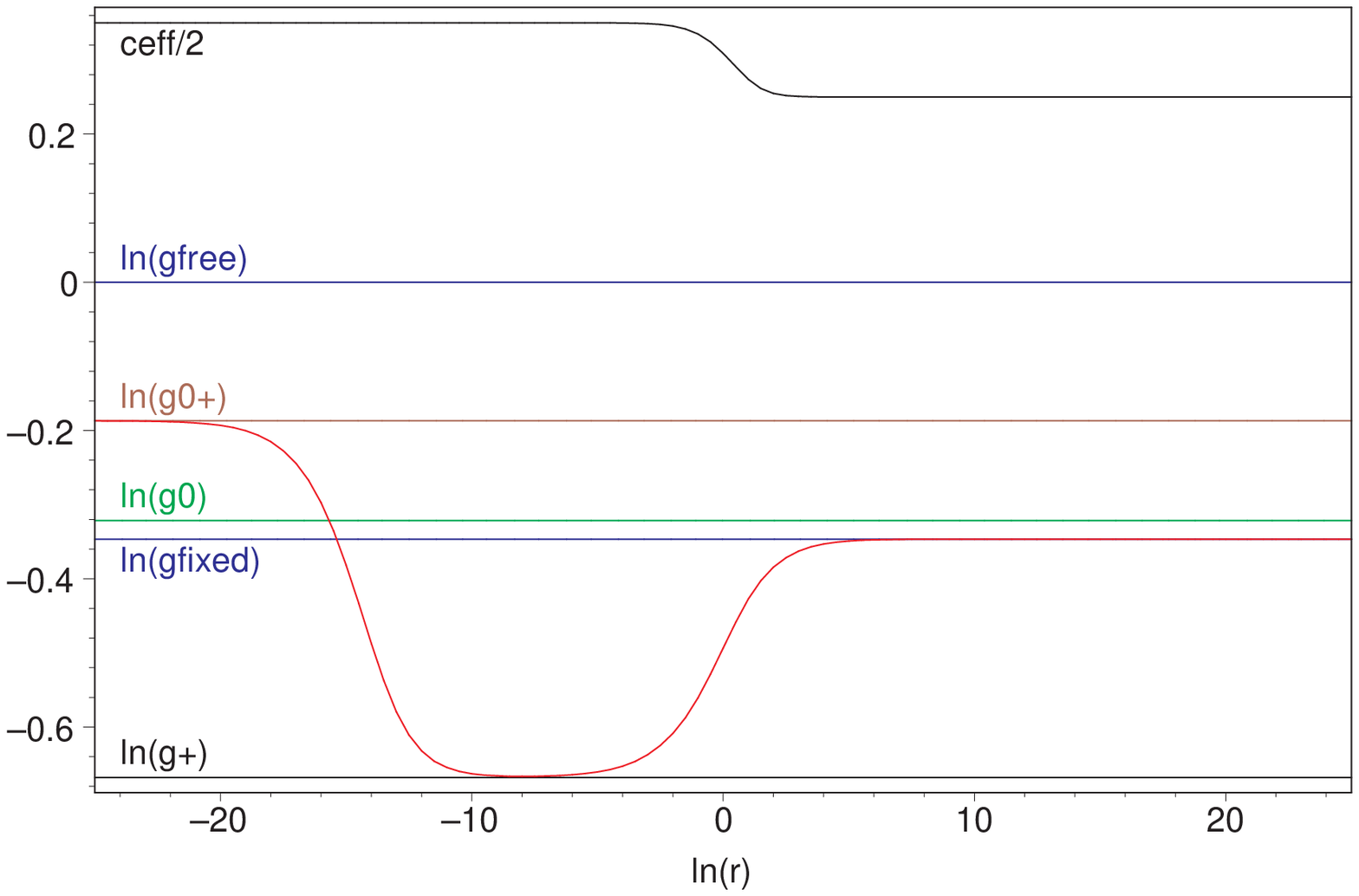}
\\[2pt]
\parbox{0.8\linewidth}{
\small Figure \protect\ref{gflowp15}: 
The exact $g$-function flow for $\theta_b=15$\,. Labelling as 
for figure \ref{gflowm15}.
}
\end{array}
\]
\refstepcounter{figure}\protect\label{gflowp15}%

The natural interpretation of the plot for $\theta_b=-15$ is that the
corresponding renormalisation group flow 
starts with a pure-boundary transition at 
$\ln r\approx -15$ from
the $(0+)|_{\rm tricritical}$ boundary to the vicinity of the
$(0)|_{\rm tricritical}$ boundary, with the
bulk remaining near to the tricritical Ising fixed point, then
undergoes a bulk-and-boundary transition with the bulk flowing from
tricritical Ising to Ising while the boundary moves from the
neighbourhood of $(0)|_{\rm
tricritical}$ to the neighbourhood of $(f)|_{\rm Ising}$, before
finally making a further boundary transition, at $\ln r\approx 15$,
to $(+)|_{\rm Ising}$.
For $\theta_b=0$, there is a single combined bulk-and-boundary
transition, from $(0+)|_{\rm tricrit}$ to $(+)|_{\rm Ising}$, at $\ln
r\approx 0$. The
absence of an independent boundary transition suggests that this
case corresponds to the boundary perturbation being zero, and we will 
give further evidence for this claim in the next section.
Finally, for $\theta_b=15$
there is a pure-boundary transition at 
$\ln r\approx -15$, from $(0+)_{\rm tricrit}$ to the neighbourhood of
$(+)_{\rm tricrit}$, followed by a bulk-and-boundary transition to
$(+)_{\rm Ising}$ at $\ln r\approx 0$, and no further transitions.

\[
\begin{array}{c}
\includegraphics[height=0.2\linewidth]{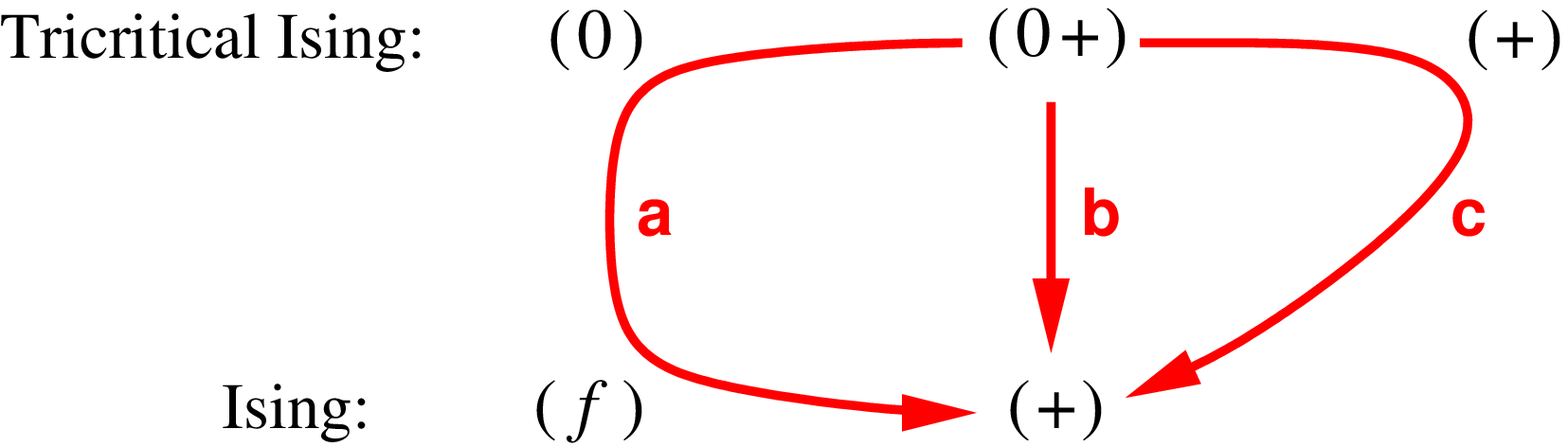}
\\[14pt]
\parbox{0.8\linewidth}{
\small Figure \protect\ref{TIM_IMflows1}: 
Combined bulk and boundary flows predicted by (\ref{gformula}),
(\ref{g0}) and (\ref{gb}).
Renormalisation group fixed points are labelled by their conformal
boundary conditions, with the bulk theory for the upper row being
the tricritical Ising model, while for
the lower it is the critical Ising model. The flows marked
{\fontsize{9.5}{0}\usefont{T1}{phv}{m}{n}a},
{\fontsize{9.5}{0}\usefont{T1}{phv}{m}{n}b} and
{\fontsize{9.5}{0}\usefont{T1}{phv}{m}{n}c} correspond to
$\theta_b=-15$, $0$ and $15$ respectively.
}
\end{array}
\]
\refstepcounter{figure}\protect\label{TIM_IMflows1}%

These results combine to give the picture sketched in
figure \ref{TIM_IMflows1}, which matches the predictions made on the
basis of large-$p$ perturbative calculations in
\cite{Fredenhagen:2009tn}. Furthermore, taking the limits 
$\theta_b\to -\infty$
and $\theta_b\to +\infty$ shows that in addition to the
$\theta_b=0$ flow
from $(0+)|_{\rm tricrit}$ to $(+)|_{\rm Ising}$, there should 
be bulk-induced flows from $(0)|_{\rm tricritical}$ to
$(f)|_{\rm Ising}$, and from $(+)|_{\rm tricritical}$ to
$(+)|_{\rm Ising}$; as explained in
\cite{Fredenhagen:2009tn}, these claims match the results of 
\cite{Pearce:2003km}.

One caveat, though: as mentioned above, strictly speaking our 
results cannot distinguish between $(+)$ and $(-)$, nor between $(0+)$
and $(-0)$, as the $g$-functions do not distinguish between these
pairs of boundary conditions. Physically it is clear that the picture
given in figure \ref{TIM_IMflows1}, and its image under a global swap
of $+$ for $-$, must be correct, but to 
resolve the issue within the context of exact $g$-function flows
alone, one would have to track the evolution of the inner products of
states other than the ground state with the boundary state. We expect
that this will be possible using pseudoenergies which solve 
excited-state TBA equations \cite{Bazhanov:1996aq,Dorey:1996re}, but 
we shall leave the further
exploration of this issue to future work.

Finally, we need a proposal for the off-critical deformations of
the $\ZZ_2$-symmetric $\phi_{13}$ flows which run from
$(d)$ up to $(+)\&(-)$ and down to $(0)$ in 
figure \ref{TIMflows}. We claim that these flows are captured by
replacing the formula (\ref{gb}) for $\ln g_b(r)$ by
\eq
\ln g_b(r)= \int_{\RR} \,
 (\phi_b(\theta)  - \phi(2 \theta) )
 \ln(1+ e^{-\varepsilon(\theta)})\, d \theta\,.
\label{gbb}
\en
In other words, we simply add $\frac{1}{2}\ln 2=0.3465\dots$ 
to the logarithm of the previous exact $g$-function. 
The graphs in figures \ref{gflowm15}, \ref{gflowp00} 
and \ref{gflowp15} are then shifted upwards by this constant, and the
transitions occur at the same values of $r$ as before, but between
a different set of conformal boundary conditions, as summarised in
figure \ref{TIM_IMflows2}. Again, this matches the extrapolation of
the predictions of \cite{Fredenhagen:2009tn} down to $p=4$.

\[
\begin{array}{c}
\includegraphics[height=0.2\linewidth]{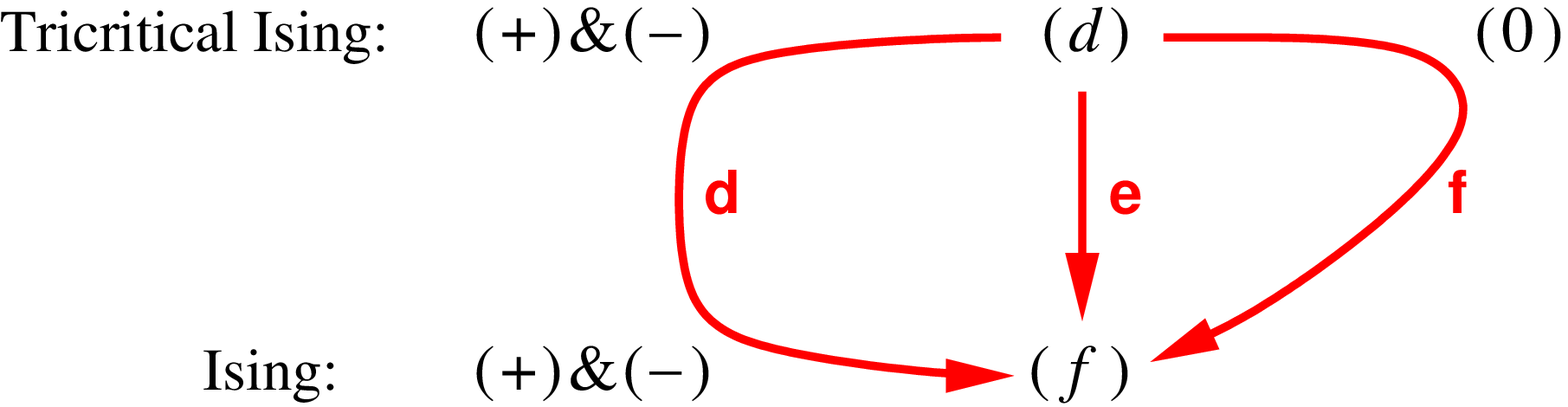}
\\[14pt]
\parbox{0.8\linewidth}{
\small Figure \protect\ref{TIM_IMflows2}: 
Combined bulk and boundary flows predicted by (\ref{gformula}),
(\ref{g0}) and (\ref{gbb}). The labelling convention for
renormalisation group fixed points is as for figure~\ref{TIM_IMflows1}.
The flows marked
{\fontsize{9.5}{0}\usefont{T1}{phv}{m}{n}d},
{\fontsize{9.5}{0}\usefont{T1}{phv}{m}{n}e} and
{\fontsize{9.5}{0}\usefont{T1}{phv}{m}{n}f} correspond to
$\theta_b=-15$, $0$ and $15$ respectively.
}
\end{array}
\]
\refstepcounter{figure}\protect\label{TIM_IMflows2}%

\resection{Exact and numerical tests of the proposal}
\label{checks}
\subsection{Exact limiting values of the g-function}
\label{exactlim}
We first deal with the universal factor $\ln g\phup_0(r)$ defined by
equation (\ref{g0}).  From (\ref{intform}),
$\int_{\RR}\phi(\theta)\,d\theta=1/2$\,,
and so
\eq
\int_{\RR^{2j-1}} d \theta_1 \dots
d \theta_{2j{-}1}\,
\phi(\theta_1+\theta_2)\phi(\theta_2+\theta_3)\dots
\phi(\theta_{2j{-}1}+\theta_{1})=\frac{1}{2^{2j}}
\label{i1}
\en
with the product of the kernel functions $\phi$
tending to zero exponentially outside a region of
order $1$ about the origin. In the infrared, only this latter
property is needed: from (\ref{IRb}), each factor
$1/(1{+}\ep(\theta_i))\to 0$
for $\theta_i\gg -\ln(r)$. In particular this holds in the
neighbourhood of the origin where the product of the kernel functions 
is significantly different from zero. Hence
all terms in the series (\ref{g0}) tend
to zero as $r\to\infty$, and $g\phup_0(r)\to 0$.

In the ultraviolet, via (\ref{UVb}), the
pseudoenergies tend to constants in the central region 
$-\ln(1/r)\ll \theta \ll\ln(1/r)$, with
$e^{-\ep(\theta)}\to (1{+}\sqrt{5})/2$\,, $1/(1{+}e^{\ep(\theta)})\to
(\sqrt{5}{-}1)/2$. Combined with (\ref{i1}), this implies
\eq
\int_{\RR^{2j-1}} \frac{d \theta_1}{1+e^{\ep(\theta_1)}} \dots
\frac{d \theta_{2j{-}1} }%
{ 1+e^{\ep(\theta_{2j{-}1})}}\,
\phi(\theta_1+\theta_2)\phi(\theta_2+\theta_3)\dots
\phi(\theta_{2j{-}1}+\theta_{1})\to
\frac{1}{2}x^{2j-1}
\en
where $x=(\sqrt{5}{-}1)/4$, and so
\eq
\lim_{r\to 0}\ln g\phup_0(r)=
\frac{1}{2}\sum_{j=1}^{\infty}\frac{1}{2j{-}1}\,x^{2j-1}=
\frac{1}{4}\ln\left(\frac{1{+}x}{1{-}x}\right)=
\frac{1}{4}\ln\left(1+\frac{2}{\sqrt{5}}\right)=
0.15972912974\dots\,.
\en

\medskip

The boundary condition dependent piece can be
treated by rewriting (\ref{gb}) so as to split $\ln g\phup_b(r)$
into three parts: a constant $\ln g\phup_{b1}$, a
parameter-independent piece $\ln g\phup_{b2}(r)$, and a
$\theta_b$-dependent piece $\ln g\phup_{b3}(r,\theta_b)$:
\eq
\ln g\phup_b(r)=\ln g\phup_{b1}+\ln g\phup_{b2}(r)+
\ln g\phup_{b3}(r,\theta_b)
\en
where
\begin{align}
\ln g\phup_{b1}&=-\frac{1}{2}\ln 2\,, \label{gb1}\\[3pt]
\ln g\phup_{b2}(r)&=
\int_{\RR}(\phi_{(3/4)}(\theta)-\phi(2\theta))L(\theta)\,d\theta\,,
\label{gb2}\\[3pt]
\ln g\phup_{b3}(r,\theta_b)&=
-\!\int_{\RR}\phi_{(1/2)}(\theta{-}\theta_b)L(\theta)\,d\theta\,.
\label{gb3}
\end{align}
Note also, from (\ref{intform}), that
\begin{align}
\int_{\RR}(\phi_{(3/4)}(\theta)-\phi(2\theta))\,d\theta
&=-\frac{1}{2}\,,
\label{int1}\\[3pt]
-\!\int_{\RR} \phi_{(1/2)}(\theta{-}\theta_b)\,d\theta
&=\frac{1}{2}\,.
\label{int2}
\end{align}
The decay properties of $\phi_{(x)}(\theta)$ mean that
the support for the integral (\ref{int1}) is concentrated near to
$\theta=0$, while that for (\ref{int2}) is concentrated near to
$\theta=\theta_b$.  Combined with the asymptotic behaviours of
$L(\theta)$ recorded in equations (\ref{UVa}) -- (\ref{IRb}), these
results 
allow the  various limiting values of $\ln g(r)$ to be computed.

\smallskip

\noindent {\bf 1}.
In the far infrared limit $\{r\to\infty,~\theta_b~\mbox{fixed}\}$,
$\ln g\phup_0(r)\to 0$, and $L(\theta)\to 0$ in the regions 
where the
integrals (\ref{gb2}) and (\ref{gb3}) can receive contributions, so
\eq
\ln g(r)\to \ln g\phup_{b1}=-\fract{1}{2}\ln 2=
\ln g\phup_{(+)}\big|_{\rm Ising} \,.
\label{lim1}
\en

\smallskip

\noindent {\bf 2}.
In the far ultraviolet limit $\{r\to 0,~\theta_b~\mbox{fixed}\}$,
$L(\theta)$ acquires a constant value in the
whole region where the integrands in (\ref{gb2}) and (\ref{gb3}) are
significantly different from zero. The integrals (\ref{gb2}) and
(\ref{gb3}) therefore cancel in the limit, and 
\eq
\ln
g(r)\to \ln g\phup_0(0)+\ln g\phup_{b1}=
\fract{1}{4}\ln(1+\fract{2}{\sqrt{5}}) -\fract{1}{2}\ln 2
=\fract{1}{4}\ln(\fract{1}{4}+\fract{1}{2\sqrt{5}})
=\ln g\phup_{(0+)}\big|_{\rm tricrit} \,.
\label{lim2}
\en

\smallskip

\noindent {\bf 3}.
If $\theta_b\ll 0$ and $\theta_b\ll \ln r \ll 0$, then 
for $\theta\approx\theta_b$, $L(\theta)\approx \ln 2$ 
from (\ref{UVa}), while 
for $\theta\approx 0$, $L(\theta)\approx \ln((3{+}\sqrt{5})/2)$ 
from (\ref{UVb}). Hence $\ln g\phup_{b3}\approx \frac{1}{2}\ln 2$ and 
$\ln g\phup_{b2}\approx -\frac{1}{2}\ln((3{+}\sqrt{5})/2)$, and
\eq
\ln g(r)\to \fract{1}{4}\ln(1{+}\fract{2}{\sqrt{5}})
 -\fract{1}{2}\ln
2-\fract{1}{2}\ln((3{+}\sqrt{5})/2)+\fract{1}{2}\ln 2
=\ln g\phup_{(0)}\big|_{\rm tricrit} \,.
\label{lim3}
\en

\smallskip

\noindent {\bf 4}.
If $\theta_b\ll 0$ and $0\ll \ln r \ll -\theta_b$, then 
for $\theta\approx\theta_b$, $L(\theta)\approx \ln 2$ from
(\ref{IRa}), while for $\theta\approx 0$, $L(\theta)\approx 0$ from
(\ref{IRb}). Hence $\ln g\phup_{b3}\approx \frac{1}{2}\ln 2$ and
$\ln g\phup_{b2}\approx 0$, and
\eq
\ln g(r)\to 
 -\fract{1}{2}\ln 2+\fract{1}{2}\ln 2=0
=\ln g\phup_{(f)}\big|_{\rm Ising} \,.
\label{lim4}
\en

\smallskip

\noindent {\bf 5}.
If $\theta_b\gg 0$ and $-\theta_b\ll \ln r \ll 0$, then 
for $\theta\approx 0$, $L(\theta)\approx \ln((3{+}\sqrt{5})/2)$
from (\ref{UVb}), while for $\theta\approx\theta_b$,
$L(\theta)\approx 0$ from (\ref{UVc}). Hence $\ln g\phup_{b2}\approx 
-\frac{1}{2}\ln((3{+}\sqrt{5})/2)$ and  $\ln g\phup_{b3}\approx 0$,
and
\eq
\ln g(r)\to \fract{1}{4}\ln(1{+}\fract{2}{\sqrt{5}})
 -\fract{1}{2}\ln
2-\fract{1}{2}\ln((3{+}\sqrt{5})/2)
=\ln g\phup_{(+)}\big|_{\rm tricrit} \,.
\label{lim5}
\en

\medskip

To make the statements of {\bf 3}, {\bf 4} and {\bf 5} 
precise, they should be considered as 
double-scaling limits: for example, for {\bf 3} one could fix
two constants
$\bar\theta_b$ and $\bar r$ with $\bar\theta_b<\ln\bar r<0$, and
set $r=\bar r^{\rho}$, $\theta_b=\rho\bar\theta_b$\,; then
(\ref{lim3}) holds in the limit $\rho\to\infty$.
If instead
$\theta_b$ is kept fixed and $r$ is varied from $0$ to $\infty$,
then the case $\theta_b\ll 0$, figure~\ref{gflowm15}, is covered by 
{\bf 1}, {\bf 3}, {\bf 4}, {\bf 2}; the case $\theta\approx
0$, figure~\ref{gflowp00}, by {\bf 1}, {\bf 2}; and the case
$\theta_b\gg 0$,
figure~\ref{gflowp15}, by {\bf 1}, {\bf 5}, {\bf 2}. 
We have thus confirmed analytically the previously-observed 
numerical results, and justified that our conjectured equations 
are indeed consistent with the flow patterns
depicted in figures \ref{TIM_IMflows1} and \ref{TIM_IMflows2}.

\subsection{Comparisons with conformal perturbation theory}
The bulk perturbation which induces the flow from the tricritical
Ising model to the Ising model corresponds to the addition of a term
$\lambda\int\phi_{13}(x,\bar x)\,d^2x$ to the action of the
tricritical model, where the bulk
coupling $\lambda$
has dimension $(\mbox{mass})^{4/5}$. 
If the unperturbed conformal
boundary condition $(\alpha)$ supports the boundary
field $\phi_{13}(x)$, the addition of a boundary perturbation 
$\mu\int \phi_{13}(x)\,dx$ can also be considered,
where $\mu$ is the boundary coupling, with dimension
$(\mbox{mass})^{2/5}$. (For the tricritical Ising conformal boundary 
conditions featured on figure
\ref{TIMflows}, $(-0)$, $(d)$ and $(0+)$ do support this
field, while $(-)$, $(0)$, $(+)$ and $(+)\&(-1)$ do not.)
A $g$-function as evaluated in conformal perturbation theory should
therefore have the expansion
\eq
\ln{\cal G}(\lambda,\mu,R)=
\sum_{m,n=0}^{\infty}c^{(\alpha)}_{m,n}(\mu R^{2/5})^m(\lambda
R^{4/5})^n\,.
\label{CPTseries}
\en
In general this is a regular series in powers of $R^{2/5}$, 
reducing 
to a series in $R^{4/5}$
when $\mu=0$.  At large $R$, the function defined by 
(\ref{CPTseries}) will typically develop a linear behaviour,
with $\ln{\cal G}(h,\lambda,R)\sim -fMR$ where $f$ is a
free energy per unit length, which we choose to measure in units of
the inverse crossover scale $M$. Our exact equations, by 
contrast, yield 
`subtracted' $g$-functions from which this term is absent in the
infrared, and instead reappears as an irregular term in the
ultraviolet \cite{Zamolodchikov:1989cf,Dorey:1999cj}. They are also
expressed in terms of $M$ and the boundary
parameter $\theta_b$, rather than $\lambda$ and $\mu$. The relation 
between $\lambda$ and $M$ is known
\cite{Zamolodchikov:1991vx,Zamolodchikov:1995xk}:
\eq
\lambda=\kappa M^{4/5}\,,\quad
\kappa=
\frac{1}{2\sqrt{2}(3\pi)^{1/5}} 
\sqrt{\frac{\Gamma(7/10)}{\Gamma(3/10)}}
=0.14869551611\dots
\label{kappaM}
\en
and on dimensional grounds it must be possible to write $\mu$ as
\eq
\mu=\nu(\theta_b) M^{2/5}
\en
where $\nu$ is some dimensionless function of $\theta_b$.
The $g$-function defined by (\ref{gformula}) should thus
have the following expansion about $r\equiv\MR=0$\,:
\eq
\ln g(r,\theta_b) = \ln g\phup_0(r)+\ln g\phup_b(r) = fr + 
\sum_{m,n=0}^{\infty}c^{(\alpha)}_{m,n}(\nu r^{2/5})^m(\kappa r^{4/5})^n
\label{TBAseries}
\en
\nobreak
where $c^{(\alpha)}_{0,0}$ is equal to the logarithm 
of the conformal $g$-function $g_{(\alpha)}$
for the $(\alpha)$ boundary condition. For all values of $\theta_b$ we
expect
\eq
c^{(\alpha)}_{1,0}=0
\label{c10}
\en
since $c^{(\alpha)}_{1,0}$ is proportional to the one-point function of 
the perturbing boundary operator on a disk with the vacuum field 
at its centre \cite{Dorey:1999cj}, and vanishes in a unitary
theory such as this one \cite{Dorey:2004xk}\footnote{Note, if
$c\phup_{1,0}$ did not vanish, then the $g$-theorem, which states 
that $g$ decreases for all pure-boundary flows in unitary models
\cite{Affleck:1991tk,Friedan:2003yc}, would be violated for one or
other sign of the boundary coupling. Conversely, the non-vanishing of
$c_{0,1}$ \cite{Dorey:2005ak} is an easy way to see that the
$g$-theorem {\em can}\/ be violated when the bulk flows, even in a
unitary theory (see also \cite{Green:2007wr}).}.
Previous examples
suggest that $\ln g\phup_0(r)$ will not contribute to the irregular
term $fr$ \cite{Dorey:1999cj,Dorey:2004xk,Dorey:2005ak}\,; 
assuming that this holds true here too, 
the value of $f$ can be calculated from (\ref{gb}) as in 
\cite{Zamolodchikov:1989cf,Dorey:1999cj}, with the result
\eq
f(\theta_b)=\fract{1}{2}e^{-\theta_b} -\fract{1}{2\sqrt{2}}\,.
\label{f}
\en
Finally, the first bulk-induced
coefficient in the expansion of $\ln g(r)$ is \cite{Dorey:2005ak}
\eq
c^{(\alpha)}_{0,1}
=-\frac{B(1{-}x_{\phi},x_{\phi}/2)}{\,2(2\pi)^{1-x_{\phi}}}%
\,\frac{\vev{\phi\,|(\alpha)}}{\vev{0\,|(\alpha)}}
\label{c01}
\en
where $B(x,y)=\Gamma(x)\Gamma(y)/\Gamma(x{+}y)$ is Euler's beta
function, $\phi$ is the bulk perturbing field, $\phi_{13}$ in this
case, and $x_{\phi}$ is its scaling dimension, here equal to $6/5$.
The inner products $\vev{\phi\,|(\alpha)}$ and 
$\vev{0\,|(\alpha)}\equiv g\phup_{(\alpha)}$ can be read from 
(\ref{TIMstates}), bearing in mind that the Ishibashi states
in those formulae have been labelled by the conformal dimensions of 
their Virasoro representations, which are half the scaling dimensions 
of the corresponding bulk fields.

The bulk TBA equation
(\ref{bulkTBA}) was solved numerically for 101 evenly-spaced values of
$r^{4/5}$ running from $0.0005$ to $0.1255$, discretising the $\theta$ axis
to $1520$ points between $\theta=-50$ and $\theta=50$ and using
extended ($20$ decimal digit) precision in GNU Fortran 95. The
resulting estimates for the pseudoenergy $\ep(\theta)$ were then 
used to compute $\ln g(r)$ via (\ref{gformula}), summing the series
(\ref{g0}) for $\ln g_0(r)$ to 12 terms, and evaluating the 
$\theta_b$-dependent part $\ln g_b(r)$ from (\ref{gb}) for values of 
$\theta_b$ ranging from $-2.5$
and $2.5$. (Were accurate results to be required for a larger 
range of $\theta_b$, care would have to be taken to decrease the 
values of $r^{4/5}$ used for the
fits, to avoid their
contamination by the
intermediate plateau values of $\ln g(r)$ which appear as $|\theta_b|$
increases, as on figures \ref{gflowm15} and~\ref{gflowp15}.)

As a first check of our numerics, we made a least-squares fit of the
function $c_{\rm eff}(r)$
defined by (\ref{ceff}) to a regular expansion in powers of
$r^{4/5}$ plus a single `antibulk' term proportional to $r^2$, 
finding coefficients which matched those reported in 
\cite{Zamolodchikov:1991vx} to the full accuracy claimed there.

Then, for each value of $\theta_b$, the numerically-obtained
$\ln g(r,\theta_b)$ was fitted to a series
in $r^{2/5}$ plus a single term proportional to $r$, as in
(\ref{TBAseries}):
\eq
\ln g(r,\theta_b) = 
\sum_{k=0}^{\infty}d\phup_k(\theta_b)\, r^{2k/5}
+e(\theta_b)\,r\,.
\label{fitseries}
\en
If the match with conformal perturbation theory is to hold,
we should have
\eq
d_k(\theta_b)=\sum_{l=0}^{\lfloor k/2\rfloor}
c^{(\alpha)}_{k-2l,l}\nu^{k-2l}\kappa^l
\,,\qquad e(\theta_b)=f(\theta_b)\,.
\en
The constant term $d_0(\theta_b)$ obtained from the fits matched the value 
predicted by (\ref{lim2}), namely
$\ln g\phup_{(0+)}\big|_{\rm tricrit}$, to at least 10 digits for the
whole range of $\theta_b$. Furthermore,
$d_1(\theta_b)$ was zero to the same accuracy, in line with (\ref{c10}).
In figure \ref{lint3}a, the values of $e(\theta_b)$ found from our
fits are compared with the exact predictions from (\ref{f}); the good
agreement supports our claim that $\ln g_0(r)$ does not
contribute to this linear term. 

\[
\begin{array}{cc}
\psfrag{ff}{$\scriptstyle f$}
\psfrag{tt}{$\theta_b$}
\!\!\!\includegraphics[width=0.5\linewidth]{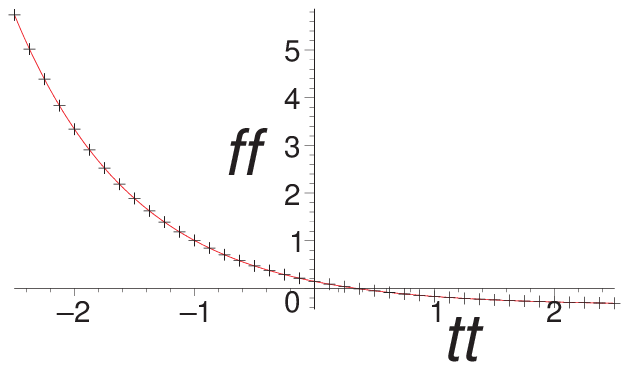}&
\psfrag{cc}{$\quad~\scriptstyle d_3$}
\psfrag{tt}{$\theta_b$}
\!\!\!\includegraphics[width=0.5\linewidth]{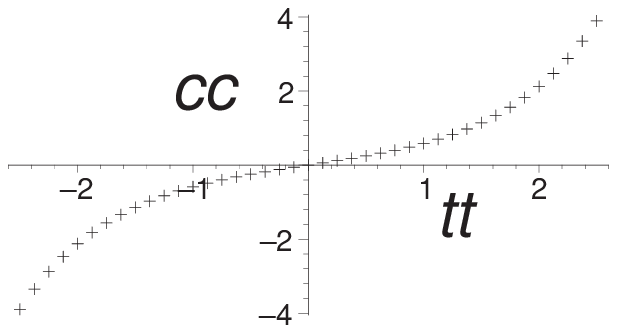}
\\[14pt]
\parbox{0.45\linewidth}{
\small Figure \protect\ref{lint3}a: A comparison of the exact
prediction (\ref{f}) for $f(\theta_b)$ (dotted line) with values
of $e(\theta_b)$ from the fit (\ref{fitseries}) (points). 
}
&
\parbox{0.45\linewidth}{
\small Figure \protect\ref{lint3}b: Estimates of $d_3(\theta_b)$ from the
fit (\ref{fitseries}). The estimate at $\theta_b=0$ is
$5.24{\times}10^{-8}$, consistent with $0$ to our numerical accuracy.
}
\end{array}
\refstepcounter{figure}\protect\label{lint3}%
\]

Next, in figure \ref{lint3}b, we show the values of $d_3(\theta_b)$.
The apparent zero of this function at $\theta_b=0$ suggests that this
point should correspond to $\mu=0$ in (\ref{TBAseries}), where {\em
all}\/ odd terms in the regular series should vanish. This is
consistent with the fit at $\theta_b=0$, which is
\begin{multline}
\!\!\!\!
\ln g(r,0) =
 -0.1868444605395000 
 +0.1464465945456
\,r
-8.429{\times}10^{-12}\,r^{2/5}\\
~~~~~
 -0.2038867755577
\,r^{4/5}
 +5.24{\times}10^{-8}\,r^{6/5}
 -0.008541571
\,r^{8/5}\qquad\qquad\\
+3.68{\times}10^{-6}\,r^2
-0.00209\,r^{12/5}
+\dots\qquad\qquad\qquad
\label{firstfit}
\end{multline}
Supposing that the odd terms are exactly zero for $\theta_b=0$, a
more-constrained fit to a regular series in powers of $r^{4/5}$ 
plus a term linear in $r$ gives the result
\begin{multline}
\!\!\!\!
\ln g(r,0) =
 -0.1868444605395363 
 +0.1464466094005
\,r
 -0.2038867770734
\,r^{4/5}\\
 -0.008541178
\,r^{8/5}
 -0.0020624
\,r^{12/5}
 +0.00151
\,r^{16/5}
  -0.0004
\,r^4
 +\dots
\label{constrainedexp}
\end{multline}
For this case, $d_2$ is known exactly, since with $\mu$ and hence
$\nu=0$, $d_2=c^{(\alpha)}_{0,1}\kappa$ and can be calculated
from (\ref{c01}) and (\ref{kappaM}). For the $(0+)$ boundary with
$\theta_b=0$ we thus have the exact predictions
\begin{align}
d_0
 &=\ln g\phup_{(0+)}=-0.1868444605395326\dots\\[3pt]
d_1&=0\\
d_2
 &=\frac{B(-1/5,3/5)\,\kappa}{\,2(2\pi)^{-1/5}\eta^3}%
=-0.2038867770751855\dots\\
d_3&=0\\[3pt]
e\,&=f(0)=0.1464466094067\dots
\end{align}
all of which are reproduced very well by the fits (\ref{firstfit}) and
(\ref{constrainedexp}).

The limits $\theta_b\to -\infty$ and $\theta_b\to +\infty$
admit similarly-simple checks. A consideration of figure
\ref{TIM_IMflows1} and the results from subsection \ref{exactlim}
shows that if these limits are taken first, keeping $r$ finite, and
$r$ is only then allowed to vary, then the resulting equations 
should describe the bulk-induced flows $(0)|_{\rm tricrit} 
\to (f)|_{\rm Ising}$ and $(+)|_{\rm tricrit}\to (+)_{\rm Ising}$ 
respectively. Neither of the UV boundary conditions 
for these flows admit a $\phi_{13}$ boundary field, so the logarithms
of their $g$-functions should have regular expansions about $r=0$ 
in powers of $r^{4/5}$, with a coefficient $d_2$ of $r^{4/5}$ that can
be predicted from (\ref{c01}).

In the limit $\theta_b\to -\infty$, 
making use of (\ref{int2}) and 
(\ref{IRa}),
equation (\ref{gb}) reduces to
\eq
\ln g_b(r)|\phup_{\theta_b=-\infty} =  \int_{\RR} \,
 (\phi_{(3/4)}(\theta)  - \phi(2 \theta) )
 L(\theta)\, d \theta 
\label{gbm}
\en
while equation (\ref{g0}) for $\ln g\phup_0(r)$ is unchanged.
With the numerical work as before, the fit to a series in powers of
$r^{2/5}$, together with a linear term, was
\begin{multline}
\!\!\!\!\!
\ln g(r)|\phup_{\theta_b=-\infty} =  
 -0.3214826953191443 
 -0.3535533993721
\,r
 -5.178{\times}10^{-12}
\,r^{2/5} \\
 +0.5337825131495
\,r^{4/5}
 +3.04{\times}10^{-8}\,r^{6/5}
 -0.01741761
\,r^{8/5} \\
 +1.98{\times}10^{-6}\,r^2
 +0.0132\,r^{12/5}
+\dots\qquad\qquad\qquad
\label{firstfit2}
\end{multline}
and the more-constrained fit to a regular series in powers of $r^{4/5}$ 
plus a linear term gave
\begin{multline}
\!\!\!\!\!\!
\ln g(r)|\phup_{\theta_b=-\infty} =  
 -0.3214826953191671 
 -0.3535533905994
\,r
 +0.5337825122412
\,r^{4/5}\\
 -0.017417394
\,r^{8/5}
 +0.0133024
\,r^{12/5}
 -0.00130
\,r^{16/5}
 -0.0008%
\,r^4
 +\dots\!\!
\label{constrainedexp2}
\end{multline}
These results can be compared with the exact predictions for the
first few coefficients for the bulk-induced flow from the $(0)$
boundary:
\begin{align}
d_0
 &=\ln g\phup_{(0)}= -0.3214826953191634\dots\\[3pt]
d_1&=0\\
d_2
 &=-\frac{B(-1/5,3/5)\,\kappa\,\eta}{\,2(2\pi)^{-1/5}}%
= 0.5337825122395085\dots\\
d_3&=0\\[3pt]
e\,&=-\fract{1}{2\sqrt{2}}= -0.3535533905932\dots
\end{align}
Again, the agreement is very good. It is also straightforward to check
analytically that this $g$-function interpolates between the desired
values -- the argument is essentially covered by cases {\bf 3} and 
{\bf 4} of the last subsection.

For the $(+)|_{\rm tricrit}\to (+)_{\rm Ising}$ flow expected to arise
in the $\theta_b\to +\infty$ limit the story is very similar. Since,
by (\ref{IRb}), $L(\theta)\to 0$ as $\theta\to+\infty$, there is
this time no modification to the constant term in (\ref{gb}), which
becomes
\eq
\ln g_b(r)|\phup_{\theta_b=+\infty} =  
-\frac{1}{2}\ln(2)+
\int_{\RR} \,
 (\phi_{(3/4)}(\theta)  - \phi(2 \theta) )
 L(\theta)\, d \theta 
\label{gbp}
\en
with the expression for $\ln g\phup_0(r)$ again unchanged. The fits
for the expansion
coefficients (apart from the constant term) are therefore
the same as before, and it is straightforward to check that these
match expectations from perturbation theory for this situation. The
same also holds for the set of $\ZZ_2$-symmetric flows predicted by 
our second proposal, (\ref{gbb}), and so we will leave the details 
to the reader.

\resection{Conclusions}
\label{conclusions}
In this paper we have proposed an extension of the exact off-critical
$g$-function equations of \cite{Dorey:2004xk,Dorey:2005ak} to cover
situations where the bulk field theory retains massless degrees of
freedom even in the far infrared, and therefore interpolates between
two different conformal field theories. While our proposals are still
conjectural, they have passed a number of non-trivial checks 
against perturbation theory, leaving us in little doubt that they are
correct. Nevertheless, a first-principles derivation from
field-theoretic considerations
would be valuable, as would an understanding via lattice models, as
obtained for the TBA equations for the bulk interpolating flow
of $c_{\rm eff}(r)$ in \cite{Pearce:1997nv}. However this may be
a hard task, and indeed previous attempts to derive exact equations 
for exact $g$-function flows have not been entirely 
successful \cite{Woynarovich:2004gc}\footnote{See
the comments at the end of section 5 of \cite{Dorey:2004xk}.}. It
would also be interesting to see whether a process of analytic
continuation could be used to relate massive and massless
$g$-functions, though again this might be delicate, given that
the functions are initially defined via a sub-leading term in the
asymptotic behaviour of cylinder partition functions.

One feature of our results is the exact equality, up to a constant
factor, of
the $g$-functions for various a-priori different flows. It seems
likely that this can be
understood through an extension of the
defect-related work of \cite{Graham:2003nc} to theories 
off-critical in the bulk\footnote{We would like to thank Daniel 
Roggenkamp for suggesting the possible relevance of this paper.} 
and it would be interesting to explore this further.

The tricritical to critical Ising flow is interesting in its own right
(see for example \cite{Zamolodchikov:1991vx,Kastor:1988ef}), but the
main reason for concentrating on this particular case
in this paper has been its relative 
simplicity, which has allowed us to make a detailed check of the
feasibility of our approach, and to illustrate the main ideas without
too many distracting complications. As mentioned in the introduction,
we expect that the general method will  be of much wider
applicability, and we hope to return to its further applications in
the future.

\bigskip
\medskip
\noindent
{\bf Acknowledgements --} 
We would like to thank
Stefan Fredenhagen, 
Matthias Gaberdiel, Daniel Roggenkamp and
G\'erard Watts for very useful discussions, encouragement and help,
and Alexei Tsvelik for pointing out reference \cite{Tsvelick:1985} to
us. We also acknowledge with gratitude the many helpful discussions
that we had on topics related to this with Aliosha Zamolodchikov.
PED and CR thank the INFN and the University of Torino for
hospitality at the start of this project, and PED thanks Sergey Frolov
and the School of Mathematics at Trinity College, Dublin for
hospitality at its end. The work
was supported in part by an STFC rolling grant, number
ST/G000433/1, and a grant from the Leverhulme Trust (PED); by
a National Research Foundation of Korea (NRF) grant 
funded by the Korean government (MEST), number 20090063066
(CR); and by an INFN grant,
number PI11~(RT).

\bigskip
\bigskip

\hrule

\bigskip

\end{document}